\documentclass[aps,preprintnumbers,nofootinbib,superscriptaddress,twocolumn,a4paper]{revtex4}

\usepackage{amsmath}
\usepackage{amsthm}
\usepackage{amssymb}
\usepackage{amstext}
\usepackage{graphicx}
\usepackage{url}
\usepackage{multirow}

\theoremstyle{plain}

\theoremstyle{definition}

\newcommand*{\comment}[1]{}

\newcommand*{\ket}[1]{| #1 \rangle}

\newcommand*{\ketbra}[2]{|#1\rangle\!\langle#2|}

\newcommand*{\id}{\openone}
\newcommand*{\ot}{\otimes}

\newcommand*{\eps}{\varepsilon}

\newcommand*{\cF}{\mathcal{F}}

\begin{document}

\title{Memory Attacks on Device-Independent Quantum Cryptography}

\author{Jonathan \surname{Barrett}}
\email{jon.barrett@rhul.ac.uk}
\affiliation{Department of Computer Science, University of Oxford, Wolfson Building, Parks Road, Oxford OX1 3QD, U.K.}
\affiliation{Department of Mathematics, Royal Holloway, University of London, Egham Hill, Egham, TW20 0EX, U.K.}
\author{Roger \surname{Colbeck}}
\email{colbeck@phys.ethz.ch}
\affiliation{Institute for Theoretical Physics, ETH Zurich, 8093
  Zurich, Switzerland.}
\affiliation{Perimeter Institute for Theoretical Physics, 31 Caroline Street North, Waterloo, ON N2L 2Y5, Canada.}
\author{Adrian \surname{Kent}}
\email{a.p.a.kent@damtp.cam.ac.uk}
\affiliation{Centre for Quantum Information and Foundations, DAMTP, Centre for
  Mathematical Sciences, University of Cambridge, Wilberforce Road,
  Cambridge, CB3 0WA, U.K.}
\affiliation{Perimeter Institute for Theoretical Physics, 31 Caroline Street North, Waterloo, ON N2L 2Y5, Canada.}

\date{$5^{\text{th}}$ August 2013}

\begin{abstract}
  Device-independent quantum cryptographic schemes aim to guarantee
  security to users based only on the output statistics of any
  components used, and without the need to verify their internal
  functionality.  Since this would protect users against untrustworthy
  or incompetent manufacturers, sabotage or device degradation, this
  idea has excited much interest, and many device-independent schemes
  have been proposed.  Here we identify a critical weakness of
  device-independent protocols that rely on public communication
  between secure laboratories.  Untrusted devices may record their
  inputs and outputs and reveal information about them via publicly
  discussed outputs during later runs.  Reusing devices thus
  compromises the security of a protocol and risks leaking secret
  data.  Possible defences include securely destroying or isolating
  used devices.  However, these are costly and often impractical.  We
  propose other more practical partial defences as well as a new
  protocol structure for device-independent quantum key distribution
  that aims to achieve composable security in the case of two parties
  using a small number of devices to repeatedly share keys with each
  another (and no other party).
\end{abstract}

\maketitle

Quantum cryptography aims to exploit the properties of quantum systems
to ensure the security of various tasks.  The best known example is
quantum key distribution (QKD), which can enable two parties to share
a secret random string and thus exchange messages secure against
eavesdropping, and we mostly focus on this task for concreteness.
While all classical key distribution protocols rely for their security
on assumed limitations on an eavesdropper's computational power, the
advantage of quantum key distribution protocols
(e.g.~\cite{BB84,Ekert}) is that they are provably secure against an
arbitrarily powerful eavesdropper, even in the presence of realistic
levels of losses and errors~\cite{Renner}.  However, the security
proofs require that quantum devices function according to particular
specifications.  Any deviation -- which might arise from a malicious
or incompetent manufacturer, or through sabotage or degradation -- can
introduce exploitable security flaws (see e.g.~\cite{GLLSKM} for
practical illustrations).

The possibility of quantum devices with deliberately concealed flaws,
introduced by an untrustworthy manufacturer or saboteur, is
particularly concerning, since (i)~it is easy to design quantum
devices that appear to be following a secure protocol but are actually
completely insecure\footnote{In BB84~\cite{BB84}, for example, a
  malicious state creation device could be programmed to secretly send
  the basis used for the encoding in an additional degree of
  freedom.}, and (ii)~there is no general technique for identifying
all possible security loopholes in standard quantum cryptography
devices.  This has led to much interest in device-independent quantum
protocols, which aim to guarantee security \emph{on the fly} by
testing the device
outputs~\cite{MayersYao,BHK,agm,SGBMPA,abgmps,MRC,HR,MPA,ColbeckThesis,PAMBMMOHLMM,CK2}:
no specification of their internal functionality is required.

Known provably secure schemes for device-independent quantum key
distribution are inefficient, as they require either independent
isolated devices for each entangled pair to ensure device-independent
security~\cite{BHK,BKP,MRC,HR,MPA}, or a large number of entangled
pairs to generate a short key~\cite{BHK,BKP,bcktwo}.  Finding an
efficient secure device-independent quantum key distribution scheme
using two (or few) devices has remained an open theoretical challenge.
Nonetheless, in the absence of tight theoretical bounds on the scope
for device-independent quantum cryptography, progress to date has
encouraged optimism (e.g.~\cite{ekertlessreal}) about the prospects
for device-independent QKD as a practical technology, as well as for
device-independent quantum randomness
expansion~\cite{ColbeckThesis,CK2,PAMBMMOHLMM} and other applications
of device-independent quantum cryptography (e.g.~\cite{scakpm}).

However, one key question has been generally neglected in work to date
on device-independent quantum cryptography, namely what happens if and
when devices are reused.  Specifically, are device-reusing protocols
\emph{composable} -- i.e.\ do individually secure protocols of this
type remain secure when combined?  It is clear that reuse of untrusted
devices cannot be \emph{universally composable}, i.e.\ such devices
cannot be securely reused for completely general purposes (in
particular, if they have memory, they must be kept secure after the
protocol).  However, for device-independent quantum cryptography to
have significant practical value, one would hope that devices can at
least be reused for the same purpose.  For example one would like to
be able to implement a QKD protocol many times, perhaps with different
parties each time, with a guarantee that all the generated keys can be
securely used in an arbitrary environment so long as the devices are
kept secure.  We focus on this type of composability here.

We describe a new type of attack that highlights pitfalls in producing
protocols that are composable (in the above sense) with
device-independent security for reusable devices, and show that for
all known protocols such composability fails in the strong sense that
purportedly secret data become completely insecure.  The leaks do not
exploit new side channels (which proficient users are assumed to
block), but instead occur through the device choosing its outputs as
part of a later protocol.

To illustrate this, consider a device-independent scheme that allows
two users (Alice and Bob) to generate and share a purportedly secure
cryptographic key.  A malicious manufacturer (Eve) can design devices
so that they record and store all their inputs and outputs.  A well
designed device-independent protocol can prevent the devices from
leaking information about the generated key \emph{during that
  protocol}.  However, \emph{when they are reused}, the devices can
make their outputs in later runs depend on the inputs and outputs of
earlier runs, and, if the protocol requires Alice and Bob to publicly
exchange at least some information about these later outputs (as all
existing protocols do), this can leak information about the original
key to Eve.  Moreover, in many existing protocols, such leaks can be
surreptitiously hidden in the noise, hence allowing the devices to
operate indefinitely like hidden spies, apparently complying with
security tests, and producing only data in the form the protocols
require, but nonetheless actually eventually leaking all the
purportedly secure data.

We stress that our results certainly do not imply that quantum key
distribution {\it per se} is insecure or impractical.  In particular,
our attacks do not apply to standard QKD protocols in which the
devices' properties are fully trusted, nor if the devices are trusted
to be memoryless (but otherwise untrusted), nor necessarily to
protocols relying on some other type of partially trusted devices.
Our target is the possibility of (full) device-independent quantum
cryptographic security, applicable to users who purchase devices from
a potentially sophisticated adversarial supplier and rely on no
assumption about the devices' internal workings.

The attacks we present raise new issues of composability and point
towards the need for new protocol designs.  We discuss some
countermeasures to our attacks that appear effective in the restricted
but relevant scenario where two users only ever use their devices for
QKD exchanges with one another, and propose a new type of protocol
that aims to achieve security in this scenario while allowing device
reuse.  Even with these countermeasures, however, we show that
security of a key generated with Bob can be compromised if Alice uses
the same device for key generation with an additional party.  This
appears to be a generic problem against which we see no complete
defence.

Although we focus on device-independent QKD for most
of this work, our attacks also apply to other device-independent
quantum cryptographic tasks.  The case of randomness expansion is
detailed in Appendix~\ref{app:rand}.

\bigskip

\noindent\emph{Cryptographic scenario.}|We use the standard
cryptographic scenario for key distribution between Alice and Bob,
each of whom has a secure laboratory.  These laboratories may be
partitioned into secure sub-laboratories, and we assume Alice and Bob
can prevent communication between their sub-laboratories as well as
between their labs and the outside world, except as authorized by the
protocol.  The setup of these laboratories is as follows.  Each party
has a trusted private random string, a trusted classical computer and
access to two channels connecting them.  The first channel is an
insecure quantum channel.  Any data sent down this can be intercepted
and modified by Eve, who is assumed to know the protocol.  The second
is an authenticated classical channel which Eve can listen to but
cannot impersonate; in efficient QKD protocols this is typically
implemented by using some key bits to authenticate communications over
a public channel.  Each party also uses a sub-laboratory to isolate
each of the untrusted devices being used for today's protocol.  They
can connect them to the insecure quantum channel, as desired, and this
connection can be closed thereafter.  They can also interact with each
device classically, supplying inputs (chosen using the trusted private
string) and receiving outputs, without any other information flowing
into or out of the secure sub-laboratory.

As mentioned before, existing device-independent QKD protocols that
have been proven unconditionally secure~\cite{BHK,HR,MPA} require
separate devices for each measurement performed by Alice and Bob with
no possibility of signalling between these devices\footnote{Within the
  scenario described above, this could be achieved by placing each
  device in its own sub-laboratory.}, or are inefficient~\cite{bcktwo}
(in terms of the amount of key per entangled pair).  For practical
device-independent QKD, we would like to remove both of these
disadvantages and have an efficient scheme needing a small number of
devices.

Since the protocols in~\cite{HR,MPA} can tolerate reasonable levels of
noise and are reasonably efficient, we look first at implementations
of protocols taking the form of those in~\cite{HR,MPA}, except that
Alice and Bob use one measurement device each, i.e., Alice (Bob) uses
the same device to perform each of her (his) measurements.  We call
these \emph{two-device} protocols (Bob also has a separate isolated
source device: see below).  The memory of a device can then act as a
signal from earlier to later measurements, hence the security proofs
of~\cite{HR,MPA} do not apply (see also~\cite{HRW} where a different
two-device setup is discussed). It is an open question whether a
secure key can be efficiently generated by a protocol of this type in
this scenario. Here we demonstrate that, even if a key can be securely
generated, repeat implementations of the protocol using the same
devices can render an earlier generated key insecure.

\bigskip

\begin{table}
\begin{enumerate}
\item \label{step:1} Entangled quantum states used in the protocol are
  generated by a device Bob holds (which is separate and kept isolated
  from his measurement device) and then shared over an insecure
  quantum channel with Alice's device.  Bob feeds his half of each
  state to his measurement device.  Once the states are received, the
  quantum channel is closed.
\item \label{step:raw} Alice and Bob each pick a random input $A_i$
  and $B_i$ to their device, ensuring they receive an output bit
  ($X_i$ and $Y_i$ respectively) before making the next input (so that
  the $i$-th output cannot depend on future inputs).  They repeat this
  $M$ times.
\item Either Alice or Bob (or both) publicly announces their
  measurement choices, and the relevant party checks that they had a
  sufficient number of suitable input combinations for the protocol.
  If not, they abort.
\item \label{step:sift} (\emph{Sifting}.) Some output pairs may be
  discarded according to some public protocol.
\item \label{step:param_est} (\emph{Parameter estimation}.) Alice
  randomly and independently decides whether to announce each
  remaining bit to Bob, doing so with probability $\mu$ (where $M\mu
  \gg 1$).  Bob uses the communicated bits and his corresponding
  outputs to compute some test function, and aborts if it lies outside
  a desired range. (For example, Bob might compute the CHSH
  value~\cite{CHSH} of the announced data, and abort if it is below
  $2.5$.)
\item (\emph{Error correction}.) \label{step:EC} Alice and Bob perform
  error correction using public discussion, in order to (with high
  probability) generate identical strings.  Eve learns the error
  correction function Alice applies to her string.
\item \label{step:PA} (\emph{Privacy amplification.}) Alice and Bob
  publicly perform privacy amplification~\cite{BBR}, producing a
  shorter shared string about which Eve has virtually no information.
  Eve similarly learns the privacy amplification function they apply
  to their error-corrected strings.
\end{enumerate}
\caption{{\bf Generic structure of the protocols we consider.}
  Although this structure is potentially restrictive, most protocols
  to date are of this form (we discuss modifications later).  Note
  that we do not need to specify the precise sub-protocols used for
  error correction or privacy amplification.  For an additional
  remark, see Part~I of the Appendix}
\label{tab:1}
\end{table}


\noindent\emph{Attacks on two-device protocols.}|Consider a QKD
protocol with the standard structure shown in Table~\ref{tab:1}.  We
imagine a scenario in which a protocol of this type is run on day~$1$,
generating a secure key for Alice and Bob, while informing Eve of the
functions used by Alice for error correction and privacy amplification
(for simplicity we assume the protocol has no sifting procedure
(Step~\ref{step:sift})).  The protocol is then rerun on day~$2$, to
generate a second key, using the same devices.  Eve can instruct the
devices to proceed as follows.  On day~$1$, they follow the protocol
honestly.  However, they keep hidden records of all the raw bits they
generate during the protocol.  At the end of day~$1$, Eve knows the
error correction and privacy amplification functions used by Alice and
Bob to generate the secure key.

On day~$2$, since Eve has access to the insecure quantum channel over
which the new quantum states are distributed, she can surreptitiously
modulate these quantum states to carry new classical instructions to
the device in Alice's lab, for example using additional degrees of
freedom in the states.  These instructions tell the device the error
correction and privacy amplification functions used on day~$1$,
allowing it to compute the secret key generated on day~$1$.  They also
tell the device to deviate from the honest protocol for randomly
selected inputs, by producing as outputs specified bits from this
secret key.  (For example, ``for input~$17$, give day~1's key bit~$5$
as output''.)  If any of these selected outputs are among those
announced in Step~\ref{step:param_est}, Eve learns the corresponding
bits of day~$1$'s secret key.  We call this type of attack, in which
Eve attempts to gain information from the classical messages sent in
Step~\ref{step:param_est}, a \emph{parameter estimation attack}.

If she follows this cheating strategy for $N\mu^{-1}<M$
input bits, Eve is likely to learn roughly $N$ bits of day~$1$'s
secret key.  Moreover, only the roughly $N$ 
output pairs from this set that are publicly compared give Alice and
Bob statistical information about Eve's cheating.  Alice and Bob
cannot a priori identify these cheating output pairs among the
$\approx \mu M$ they compare.  Thus, if the tolerable noise level is
comparable to $N \mu^{-1} M^{-1}$, Eve can (with high probability)
mask her cheating as noise.  (Note that in unconditional security
proofs it is generally assumed that eavesdropping is the cause of all
noise.  Even if in practice Eve cannot reduce the noise to zero, she
can supply less noisy components than she claims and use the extra
tolerable noise to cheat).

In addition, Alice and Bob's devices each separately have the power to
cause the protocol to abort on any day of their choice.  Thus -- if
she is willing to wait long enough -- Eve can program them to
communicate some or all information about their day~$1$ key, for
instance by encoding the relevant bits as a binary integer $N=b_1
\ldots b_m$ and choosing to abort on day~$(N+2)$\footnote{In
  practice, Eve might infer a day $(N+2)$ abort from the fact that
  Alice and Bob have no secret key available on day $(N+2)$, which in
  many scenarios might detectably affect their behaviour then or
  subsequently.  Note too that she might alternatively program the
  devices to abort on every day from $(N+2)$ onwards if this made $N$
  more easily inferable in practice.}.  We call this type of attack an
\emph{abort attack}.  Note that it cannot be detected until it is too
late.

As mentioned above, some well known protocols use many independent and
isolated measurement devices.  These protocols are also vulnerable to
memory attacks, as explained in Appendix~\ref{app:multi}.

\bigskip

\noindent\emph{Modified protocols.}|We now discuss ways in which these
attacks can be partly defended against.\smallskip

\noindent\emph{Countermeasure 1.}|All quantum data and all public
communication of output data in the protocol come from one party, say
Bob.  Thus, the entangled states used in the protocol are generated by
a separate isolated device held by Bob (as in the protocol in Table~1)
and Bob (rather than Alice) sends selected output data over a public
channel in Step~\ref{step:param_est}.  If Bob's device is forever kept
isolated from incoming communication, Eve has no way of sending it
instructions to calculate and leak secret key bits from day~$1$ (or
any later day).

Existing protocols modified in this way are still insecure if reused,
however.  For example, in a modified parameter estimation attack, Eve
can pre-program Bob's device to leak raw key data from day~$1$ via
output data on subsequent days, at a low enough rate (compared to the
background noise level) that this cheating is unlikely to be detected.
If the actual noise level is lower than the level tolerated in the
protocol, and Eve knows both (a possibility Alice and Bob must allow
for), she can thereby eventually obtain all Bob's raw key data from
day~$1$, and hence the secret key.

In addition, Eve can still communicate with Alice's device, and Alice
needs to be able to make some public communication to Bob, if only to
abort the protocol.  Eve can thus obtain secret key bits from day~$1$
on a later day using an abort attack.\smallskip

\noindent\emph{Countermeasure 2.~\cite{Lev_comm}}|Encrypt the
parameter estimation information sent in Step~\ref{step:param_est}
with some initial pre-shared seed randomness.  Provided the seed
required is small compared to the size of final string generated
(which is the case in efficient QKD protocols \cite{HR,MPA}), the
protocol then performs key expansion\footnote{QKD is often referred
  to as quantum key expansion in any case, taking into account that a
  common method of authenticating the classical channel uses
  pre-shared randomness.}.  Furthermore, even if they have
insufficient initial shared key to encrypt the parameter estimation
information, Alice and Bob could communicate the parameter estimation
information unencrypted on day~$1$, but encrypt it on subsequent days
using generated key.

Note that this countermeasure is not effective against abort attacks,
which can now be used to convey all or part of their day~$1$ raw key.
This type of attack seems unavoidable in any standard cryptographic
model requiring composability and allowing arbitrarily many device
reuses if either Alice or Bob has only a single measurement device.

This countermeasure is also not effective in general cryptographic
environments involving communication with multiple users who may not
all be trustworthy.  Suppose that Alice wants to share key with Bob on
day~1, but with Charlie on day~2.  If Charlie becomes corrupted by
Eve, then, for example by hiding data in the parameter estimation, Eve
can learn about day~1's key (we call this an \emph{impostor attack}).
This attack applies in many scenarios in which users might wish to use
device-independent QKD.  For example, suppose Alice is a merchant and
Bob is a customer who needs to communicate his credit card number to
Alice via QKD to complete the sale.  The next day, Eve can pose as a
customer, carry out her own QKD exchange with Alice, and extract
information about Bob's card number without being detected.\smallskip

\noindent\emph{Countermeasure 3.}|Alternative protocols using
additional measurement devices.  Suppose Alice and Bob each have $m$
measurement devices, for some small integer $m\geq 2$.  They perform
Steps~\ref{step:1}--\ref{step:EC} of a protocol that takes the form
given in Table~\ref{tab:1} but with Countermeasures~1 and~2 applied.
They repeat these steps for each of their devices in turn, ensuring no
communication between any of them (i.e., they place each in its own
sub-laboratory).  This yields $m$ error-corrected strings.  Alice and
Bob concatenate their strings before performing privacy amplification
as in Step~\ref{step:PA}.  However, they further shorten the final
string such that it would (with near certainty) remain secure if one
of the $m$ error-corrected strings were to become known to Eve through
an abort attack. (See Table~2, and Appendix~\ref{app:PA} for more details).

This countermeasure doesn't avoid impostor attacks.  Instead, the idea
is to prevent useful abort attacks (as well as parameter estimation
attacks due to Countermeasure~2), and hence give us a secure and
composable protocol, provided the keys produced on successive days are
always between the same two users.  The information each device has
about day~1's key is limited to the raw key it produced.  Thus, if
each device is programmed to abort on a particular day that encodes
their day~1 raw key, after an abort, Eve knows one of the devices' raw
keys and has some information on the others (since she can exclude
certain possibilities based on the lack of abort by those devices so
far).  After an abort, Alice and Bob should cease to use any of their
devices unless and until such time that they no longer require that
their keys remain secret.  Intuitively, provided the set of $m$ keys
was sufficiently shortened in the privacy amplification step, Eve has
essentially no information about the day~1 secret key, which thus (we
conjecture) remains secure.\smallskip

\noindent\emph{Countermeasure 4.}|Alice and Bob share a small
initial secret key and use part of it to choose the privacy
amplification function in Step~\ref{step:PA} of the protocol, which
may then never become known to Eve.

Even in this case, Eve can pre-program Bob's measurement device to
leak raw data from day~1 on subsequent days, either via a parameter
estimation attack or via an abort attack.  While Eve cannot obtain
bits of the secret key so directly in this case, provided the protocol
is composed sufficiently many times, she can eventually obtain all the
raw key.  This means that Alice and Bob's residual security ultimately
derives only from the initial shared secret key: their QKD protocol
produces no extra permanently secure data.

\bigskip

In summary, we have shown how a malicious manufacturer who wishes to
mislead users or obtain data from them can equip devices with a memory
and use it in programming them.  The full scope of this threat seems
to have been overlooked in the literature on device-independent
quantum cryptography to date.  A task is potentially vulnerable to our
attacks if it involves secret data generated by devices and if Eve can
learn some function of the device outputs in a subsequent protocol.
Since even causing a protocol to abort communicates some information
to Eve, the class of tasks potentially affected is large indeed.  In
particular, for one of the most important applications, QKD, none of
the protocols so far proposed remain composably secure in the case
that the devices are supplied by a malicious adversary.

One can think of the problems our attacks raise as a new issue of
cryptographic composability.  One way of thinking of standard
composability is that a secure output from a protocol must still have
all the properties of an ideal secure output when combined with other
outputs from the same or other protocols.  The device-independent key
distribution protocols we have examined fail this test because the
reuse of devices can cause later outputs to depend on earlier ones.
In a sense, the underlying problem is that the {\it usage of devices}
is not composably secure.  This applies too, of course, for devices
used in different protocols: devices used for secure randomness
expansion cannot then securely be used for key distribution without
potentially compromising the generated randomness, for example.

It is worth reiterating that our attacks do not apply against
protocols where the devices are trusted to be memoryless.  Indeed,
there are schemes that are composably secure for memoryless
devices~\cite{HR,MPA}.  We also stress that our attacks do not apply
to all protocols for device-independent quantum tasks related to
cryptography.  For example, even devices with memories cannot mimic
nonlocal correlations in the absence of shared
entanglement~\cite{BCHKP,Gill}.  In addition, in applications that
require only short-lived secrets, devices may be reused once such
secrets are no longer required.  Partially secure device-independent
protocols for bit commitment and coin tossing~\cite{scakpm}, in which
the committer supplies devices to the recipient, are also immune from
our attacks, so long as the only data entering the devices come from
the committer.

Note too that, in practice the number of uses required to apply the
attacks may be very large, for example, in the case of some of the
abort attacks we described.  One can imagine a scenario in which Alice
and Bob want to carry out device-independent QKD no more than $n$
times for some fixed number $n$, each is confident in the other's
trustworthiness throughout, the devices are used for no other purpose
and are destroyed after $n$ rounds, and key generation is suspended
and the devices destroyed if a single abort occurs.  If the only
relevant information conveyed to Eve is that an abort occurs on one of
the $n$ days, she can only learn at most $\log n$ bits of information
about the raw key via an abort attack.  Hence one idea is that, using
suitable additional privacy amplification, Alice and Bob could produce
a device-independent protocol using two measurement devices that is
provably secure when restricted to no more than $n$ bilateral uses.
It would be interesting to analyse this possibility, which, along with
the protocol presented in Table~2, leads us to hold out the hope of
useful security for fully device-independent QKD, albeit in restricted
scenarios.

We have also discussed some possible defences and countermeasures
against our attacks.  A theoretically simple one is to dispose of --
i.e.\ securely destroy or isolate -- untrusted devices after a single
use (see Appendix~\ref{app:toxic}).  While this would
restore universal composability, it is clearly costly and would
severely limit the practicality of device-independent quantum
cryptography.  Another interesting possibility is to design protocols
for composable device-independent QKD guaranteed secure in more
restricted scenarios.    However, the impostor attacks described
above appear to exclude the possibility of composably secure
device-independent QKD when the devices are used to exchange
key with several parties.  

Many interesting questions remain open.  Nonetheless, the attacks we
have described merit a serious reappraisal of current protocol designs
and, in our view, of the practical scope of universally composable
quantum cryptography using completely untrusted devices.\bigskip

\noindent{\bf Added Remark:} Since the first version of this paper,
there has been new work in this area that, in part, explores
countermeasure~2 in more detail~\cite{MS}.  In addition, two new works
on device-independent QKD with only two devices have
appeared~\cite{RUV,VV2}.  Note that these do not evade the attacks we
present, but apply to the scenario where used devices are
discarded.\bigskip

\noindent\emph{Acknowledgements.}|We thank Anthony Leverrier and
Gonzalo de la Torre for~\cite{Lev_comm}, Llu\'is Masanes, Serge Massar
and Stefano Pironio for helpful comments.  JB was supported by the
EPSRC, and the CHIST-ERA DIQIP project.  RC acknowledges support from
the Swiss National Science Foundation (grants PP00P2-128455 and
20CH21-138799) and the National Centre of Competence in Research
`Quantum Science and Technology'.  AK was partially supported by a
Leverhulme Research Fellowship, a grant from the John Templeton
Foundation, and the EU Quantum Computer Science project (contract
255961).  This research is supported in part by Perimeter Institute
for Theoretical Physics.  Research at Perimeter Institute is supported
by the Government of Canada through Industry Canada and by the
Province of Ontario through the Ministry of Research and Innovation.


\appendix

\section{Separation of sources and measurement devices}

We add here one important comment about the general structure of the
generic protocol given in Table~1 of the main text.  There it was
crucial that in Step~1, in the case where Bob (rather than Eve)
supplies the states, he does so using a device that is isolated from
his measurement device.  If, on the other hand, Bob had only a single
device that both supplies states and performs measurements, then his
device can hide information about day~1's raw key in the states he
sends on day~2.  (This can be done using states of the form specified
in the protocol, masking the errors as noise as above.  Alternatively,
the data could be encoded in the timings of the signals or in quantum
degrees of freedom not used in the protocol.)

\section{Toxic device disposal}\label{app:toxic}
As noted in the main text, standard cryptographic models postulate
that the parties can create secure laboratories, within which all
operations are shielded from eavesdropping.  Device-independent
quantum cryptographic models also necessarily assume that devices
within these laboratories cannot signal to the outside -- otherwise
security is clearly impossible.  Multi-device protocols assume that
the laboratories can be divided into effectively isolated
sub-laboratories, and that devices in separate sub-laboratories cannot
communicate.  In other words, Alice and Bob must be able to build
arbitrary configurations of screening walls, which prevent
communication among Eve and any of her devices, and allow only
communications specified by Alice and Bob.

Given this, there is no problem {\it in principle} in defining
protocols which prescribe that devices must be permanently isolated:
the devices simply need to be left indefinitely in a screened
sub-laboratory.  While this could be detached from the main working
laboratory, it must be protected indefinitely: screening wall material
and secure space thus become consumed resources.  
And indeed in some situations, it may be more efficient to isolate
devices, rather than securely destroy them, since devices can be
reused once the secrets they know have become public by other means.
For example, one may wish to securely communicate the result of an
election before announcing it, but once it is public, the devices used
for this secure communication could be safely reused.

The alternative, securely destroying devices and then eliminating them
from the laboratory, preserves laboratory space but raises new
security issues: consider, for example, the problems in disposing of a
device programmed to change its chemical composition depending on its
output bit.

That said, no doubt there are pretty secure ways of destroying
devices, and no doubt devices could be securely isolated for long
periods.  However, the costs and problems involved, together with the
costs of renewing devices, make us query whether these are really
viable paths for practical device-independent quantum cryptography.

\section{Privacy Amplification}\label{app:PA}
Here we briefly outline the important features of privacy
amplification, which is a key step in the protocol.  As explained in
the main text, the idea is to compress the string such that (with high
probability) an eavesdropper's knowledge is reduced to nearly zero.
This usually works as follows.  Suppose Alice and Bob share some
random string, $X$, which may be correlated with a quantum system, $E$,
held by the eavesdropper.  Alice also holds some private randomness,
$R$.  The state held by Alice and Eve then takes the form
$$\rho_{XRE}=\sum_{x,r}P_X(x)P_R(r)\ketbra{x}{x}_X\ot\ketbra{r}{r}_R\ot\rho_E^x,$$
where $\{\rho_E^x\}_x$ are normalized density operators, and
$P_R(r)=1/|R|$.  The randomness $R$ is used to choose a function
$f_R\in\cF$, where $\cF$ is some suitably chosen set, to apply to $X$
such that, even if she learns $R$, the eavesdropper's knowledge about
the final string is close to zero.  If we call the final string
$S=f_R(X)$, then Eve has no knowledge about it if the final state
takes the form $\tau_S\ot\rho_{RE}$, where $\tau_S$ is maximally
mixed on $S$.  However, we cannot usually attain such a state, and instead
measure the success of a protocol by its variation from this ideal,
measured using the \emph{trace distance}, $D$.  Denoting the final
state (after applying the function) by $\rho_{SRE}$, we are interested
in $D(\rho_{SRE},\tau_S\ot\rho_{RE})$.

Fortunately, several sets of function are known for which the above
distance can be made arbitrarily small.  Two common constructions are
those based on \emph{two-universal hash
  functions}~\cite{CW,WC,Renner,TSSR} and \emph{Trevisan's
  extractor}~\cite{Trevisan,DPVR}.  The precise details of these is
not very important for the present work (we refer the interested
reader to the references), nor is it important which we choose.
However, it is worth noting that for two-universal hash functions, the
size of the seed needs to be roughly equal to that of the final
string, while for Trevisan's extractor, this can be reduced to roughly
the logarithm of the length of the initial string (in the latter case,
this may allow it to be sent privately, if desired).

For both, the amount that the string should be compressed
is quantified by the smooth conditional min-entropy, which we now
define.  For a state $\rho_{AB}$, the non-smooth conditional
min-entropy is defined as
$$H_{\min}(A|B)_{\rho}:=\max_{\sigma_B}\sup\{\lambda\in\mathbb{R}:2^{-\lambda}\id_A\ot\sigma_B\geq\rho_{AB}\},$$
in terms of which the smooth min entropy is given by
$$H_{\min}^{\eps}(A|B)_{\rho}:=\max_{\bar{\rho}_{AB}}H_{\min}(A|B)_{\bar{\rho}}.$$
The maximization over $\bar{\rho}$ is over a set of states that
are close to $\rho_{AB}$ according to some distance measure (see, for
example,~\cite{TCR2} for a discussion).

The significance for privacy amplification can be seen as follows.
In~\cite{Renner}, it is shown that if $f$ is chosen randomly from a
set of two-universal hash functions, and applied to the raw string
$X$, as above, then for $|S|=2^t$ and any $\eps\geq 0$,
$$D(\rho_{SRE},\tau_S\ot\rho_{RE})\leq\eps+\frac{1}{2}2^{-\frac{1}{2}(H_{\min}^\eps(X|E)-t)}.$$
(An analogous statement can be made for Trevisan's
extractor~\cite{DPVR}.)  Thus, if Alice compresses her string to
length $t=H_{\min}^\eps(X|E)-\ell$, then the final state after
applying the hash function has distance $\eps+\frac{1}{2}2^{-\ell/2}$
to a state about which Eve has no knowledge.

\begin{table}
\renewcommand\thetable{2}
\begin{enumerate}
\item \label{step:1a} Entangled quantum states used in the protocol are
  generated by a device Bob holds (which is separate and kept isolated
  from his measurement devices) and then shared over an insecure
  quantum channel with Alice's first device.  Bob feeds his half of
  each state to his first measurement device.  Once the states are
  received, the quantum channel is closed.
\item \label{step:rawa} Alice and Bob each pick a random input $A_i$
  and $B_i$ to their first device, ensuring they receive an output bit
  ($X_i$ and $Y_i$ respectively) before making the next input (so that
  the $i$-th output cannot depend on future inputs).  They repeat this
  $M$ times.
\item Bob publicly announces his measurement choices, and Alice checks
  that for a sufficient number of suitable input combinations for the
  protocol.  If not, Alice aborts.
\item \label{step:sifta} (\emph{Sifting}.) Some output pairs may be
  discarded according to some protocol.
\item \label{step:param_esta} (\emph{Parameter estimation}.) Alice and
  Bob use their pre-shared key to randomly select some output pairs
  (they select only a small fraction, hence the amount of key required
  for this is small).  For each of the selected pairs, Bob encrypts his
  output and sends it to Alice.  Alice uses the communicated bits and
  her corresponding outputs to compute some test function, and aborts
  if it lies outside a desired range.
\item (\emph{Error correction}.) \label{step:ECa} Alice and Bob perform
  error correction using public discussion, in order to (with high
  probability) generate identical strings.  Eve learns the error
  correction function Alice applies to her string.
\item Alice and Bob repeat Steps~\ref{step:1a}--\ref{step:ECa} for each
  of their $m$ devices (ensuring the devices cannot communicate
  throughout)
\item \label{step:PAa} (\emph{Privacy amplification.}) Alice and Bob
  concatenate their $m$ strings and publicly perform privacy
  amplification~\cite{BBR}, producing a shorter shared string about
  which Eve has virtually no information.  In this step, the size of
  their final string is chosen such that (with high probability) it
  will remain secure even if one of the raw strings or its error
  corrected version becomes known.
\end{enumerate}
\caption{{\bf Structure of the protocol from the main text with
    modifications as in Countermeasure~3.}  For this protocol Alice
  and Bob each have $m\geq 2$ measurement devices, and Bob has one device
  for creating states.  They are all kept isolated from one another.}
\label{tab:2}
\end{table}

Turning to the QKD protocol in Table~1 of the main text, in the case
of hashing the privacy amplification procedure consists of Alice
selecting $t$ depending on the test function computed in the parameter
estimation step.  She then uses local randomness to choose a hash
function to apply to her string, and announces this to Bob, who
applies the same function to his string (since we have already
performed error correction, this string should be identical to
Alice's).  The idea is that, if $t$ is chosen appropriately, it is
virtually impossible that the parameter estimation tests pass and the
final state at the end of the protocol is not close to one for which
Eve has no knowledge about the final string.

In the modified protocol in Table~2, we expect each pair of devices to
contribute roughly the same amount of smooth min entropy to the
concatenated string.  Thus, since there are $m$ devices, in order to
tolerate the potential revelation of one of the error-corrected
strings through an abort attack, Alice should choose $t$
to be roughly $(m-1)/m$ shorter than she would otherwise.

\section{Memory attacks on multi-device QKD protocols}
\label{app:multi}
To illustrate further the generality of our attacks, we now turn to
multi-device protocols, and show how to break iterated versions of
two well known protocols.

\subsection*{Attacks on compositions of the BHK protocol}
The Barrett-Hardy-Kent (BHK) protocol~\cite{BHK} requires Alice and
Bob to share $MN^2$ pairs of systems (where $M$ and $N$ are both large
with $M \ll N$), in such a way that no measurements on any subset can
effectively signal to the others.  In a device-independent scenario,
we can think of these as black box devices supplied by Eve, containing
states also supplied by Eve.  Each device is isolated within its own
sub-laboratory of Alice's and Bob's, so that Alice and Bob have $MN^2$
secure sub-laboratories each.  The devices accept integer inputs in
the range $\{0,\ldots,N-1\}$ and produce integer outputs in the
range $\{0,1\}$.  Alice and Bob choose random independent inputs,
which they make public after obtaining all the outputs.  They also
publicly compare all their outputs except for those corresponding to
one pair randomly chosen from among those in which the inputs differ
by $\pm 1$ or $0$ modulo $N$.  If the publicly declared outputs agree
with quantum statistics for specified measurement basis choices
(corresponding to the inputs) on a singlet state, then they accept the
protocol as secure, and take the final undeclared outputs (which are
almost certainly anticorrelated) to define their shared secret bit.

The BHK protocol produces (with high probability) precisely one secret
bit: evidently, it is extremely inefficient in terms of the number of
devices required.  It also requires essentially noise-free channels
and error-free measurements.  Despite these impracticalities it
illustrates our theoretical point well.  Suppose that Alice and Bob
successfully complete a run of the BHK protocol and then (unauthorised
by BHK) decide to use the same $2M N^2$ devices to generate a second
secret bit, and ask Eve to supply a second batch of states to allow
them to do this.

Eve --- aware in advance that the devices may be reused --- can design
them to function as follows.  In the first run of the protocol, she
supplies a singlet pair to each pair of devices and the devices
function honestly, carrying out the appropriate quantum measurements
on their singlets and reporting the outcomes as their outputs.
However, they also store in memory their inputs and outputs.  In the
second run, Eve supplies a fresh batch of singlet pairs.  However, she
also supplies a hidden classical signal identifying the particular
pair of devices that generated the first secret bit.  (This signal
need go to just one of this pair of devices, and no others.)  On the
second run, the identified device produces as output the same output
that it produced on the first run (i.e. the secret bit generated, up
to a sign convention known to Eve).  All other devices function
honestly on the second run.

With probability $\frac{MN^2-1}{M N^2}$, the output from the cheating
device on the second run will be made public, thus revealing the first
secret bit to Eve.  Moreover, with probability $ 1 - \frac{3}{2N} + O
( N^{-2} )$, this cheating will not be detected by Alice and Bob's
tests, so that Eve learns the first secret bit without her cheating
even being noticed.

There are defences against this specific attack.  First, the BHK
protocol \cite{BHK} can be modified so that only outputs corresponding
to inputs differing by $\pm 1$ or $0$ are publicly shared.\footnote{As
  originally presented, the BHK protocol requires public exchange of
  all outputs except those defining the secret key bit.  This is
  unnecessary, and makes iterated implementations much more vulnerable
  to the attacks discussed here.}  While this causes Eve to wait many
rounds for the secret bit to be leaked, and increases the risk her
cheating will be detected, it leaves the iterated protocol insecure.
Second, Alice and Bob could securely destroy or isolate the devices
producing the secret key bit outputs, and reuse all their other
devices in a second implementation.  Since only the devices generating
the secret key bit have information about it, this prevents it from
being later leaked.  While effective, this last defence really
reflects the inefficiency of the BHK protocol: to illustrate this, we
turn next to a more efficient multi-device protocol.


\subsection*{Attacks on compositions of the HR protocol}

H\"anggi and Renner (HR)~\cite{HR} consider a multi-device QKD
protocol related to the Ekert~\cite{Ekert} protocol, in which Alice
and Bob randomly and independently choose one of two or three inputs
respectively for each of their devices.  If the devices are
functioning honestly, these correspond to measurements of a shared
singlet in the bases $U_0, U_1 $ (Alice) and $V_0 , V_1 , V_2$ (Bob),
defined by the following vectors and their orthogonal complements
\begin{eqnarray*}
U_1 & \leftrightarrow & \ket{0} \, , \\
V_0 & \leftrightarrow & \cos ( \pi / 8 )   \ket{0} + \sin ( \pi / 8 )
  \ket{1} \, ,  \\
U_0 , V_2 & \leftrightarrow & \cos ( \pi / 4 )   \ket{0} + \sin ( \pi / 4 )
  \ket{1} \, ,  \\
V_1 & \leftrightarrow &   \cos ( 3 \pi / 8 )   \ket{0} + \sin ( 3 \pi / 8 )
  \ket{1} \, . 
\end{eqnarray*}

The raw key on any given run is defined by the $\approx 1/6$ of the
cases in which $U_0$ and $V_2$ are chosen.  Information reconciliation
and privacy amplification proceed according to protocols of the type
described in the main text (in which the functions used are released
publicly).

Evidently, our attacks apply here too if (unauthorised by HR) the
devices are reused to generate further secret keys.  Eve can identify
the devices that generate the raw key on day~$1$, and request them to
release their key as cheating outputs on later days, gradually enough
that the cheating will be lost in the noise.  Since the information
reconciliation and privacy amplification functions were made public by
Alice, she can then obtain the secret key.  Even if she is unable to
communicate directly with the devices for a long time (because they
were pre-installed with a very large reservoir of singlets), she can
program all devices to gradually release their day~1 outputs over
subsequent days, and so can still deduce the raw and secret keys.

Alice and Bob could counter these attacks by securely destroying or
isolating all the devices that generated raw key on day~$1$ --- but
this costs them $1/6$ of their devices, and they have to apply this
strategy each time they generate a key, leaving $(5/6)^N$ of the
devices after $N$ runs, and leaving them able to generate shorter and
shorter keys.  As the length of secure key generated scales by
$(5/6)^N$ (or worse, allowing for fluctuations due to noise) on each
run, the total secret key generated is bounded by $\approx 6M$, where
$M$ is the secret key length generated on day~$1$.

Note that, as in the case of the iterated BHK protocol, all devices
that generate secret key become toxic and cannot be reused.  While the
relative efficiency of the HR protocol ensures a (much) faster secret
key rate, it also requires an equally fast device depletion rate.
This example shows that our attacks pose a generic problem for
device-independent QKD protocols of the types considered to date.

\section{Device-independent randomness expansion protocols: attacks
  and defences}\label{app:rand}
Device-independent quantum randomness expansion (DVI QRE) protocols
were introduced by two of us~\cite{ColbeckThesis,CK2}, developed
further by~\cite{PAMBMMOHLMM,FGS,VV,PM}, and there now exist schemes
with unconditional security proofs~\cite{VV}.  The cryptographic
scenario here is slightly different from that of key distribution in
that there is only one honest party, Alice.

Alice's aim is to expand an initial secret random string to a
longer one that is guaranteed secret from an eavesdropper, Eve, even
if the quantum devices and states used are supplied by Eve.  The
essential idea is that seed randomness can be used to carry out
nonlocality tests on the devices and states, within one or more secure
laboratories, in a way that guarantees (with numerical bounds) that
the outcomes generate a partially secret and random string.  Privacy
amplification can then be used to generate an essentially fully secret
random string, which (provided the tests are passed) is significantly
longer than the initial seed.

There are already known pitfalls in designing such protocols.  For example,
although one might think that carrying out a protocol in a single
secure laboratory guarantees that the initially secure seed string
remains secure, and so guarantees randomness expansion if any new
secret random data is generated, this is not the case~\cite{CK2}.
Eve's devices may be programmed to produce outputs depending on the
random seed in such a way that the length of the final secret random
string depends on the initial seed.  Protocols with this vulnerability
are not composably secure.  (To see this can be a practical problem,
note that Eve may infer the length of the generated secret random
string from its use.)

A corollary of our results is that, if one wants to reuse the devices
to generate further randomness, it is crucial to carry out DVI QRE
protocols with devices permanently held within a {\it single} secure
laboratory, avoiding any public communication of device output data at
any stage.  It is crucial too that the devices themselves are securely
isolated from classical communications and computations within the
laboratory, to prevent them from learning details of the
reconciliation and privacy amplification.

Even under these stringent conditions, our attacks still apply in
principle.  For example, consider a noise-tolerant protocol that
produces a secret random output string of variable length, depending
on the values of test functions of the device outputs (the analogue of
QKD parameter estimation for QRE) that measure how far the device
outputs deviate from ideal honest outputs.  This might seem natural
for any single run, since -- if the devices are never reused -- the
length of the provably secret random string that can be generated does
indeed depend on the value of a suitable test function.  However,
iterating such a protocol allows the devices to leak information about
(at least) their raw outputs on the first run by generating artificial
noise in later rounds, with the level of extra noise chosen to depend
suitably on the output values.  Such noise statistically affects the
length of the output random strings on later rounds.

In this way, suitably programmed devices could ultimately allow Eve to
infer all the raw outputs from the first round, given observation of
the key string lengths created in later rounds.  This makes the round
one QRE insecure, since given the raw outputs for round one, and
knowing the protocol, Eve knows all information about the output
random string for round one, except that determined by the secret
random seed.

One defence against this would be to fix a length $L$ for the random
string generated corresponding to a maximum acceptable noise level,
and then to employ the Procrustean tactic of always reducing the
string generated to length $L$, regardless of the measured noise
level.

Even then, though, unless some restriction is placed on the number of
uses, the abort attack on QKD protocols described in the main text
also applies here.  The devices have the power to cause the protocol
to abort on any round of their choice, and so -- if she is willing to
wait long enough -- Eve can program them to communicate any or all
information about their round $1$ raw outputs by choosing the round on
which they cause an abort.

We also described in the main text a moderately costly but apparently
effective defence against abort attacks on QKD protocols, in which
Alice and Bob each have several isolated devices that independently
generate raw sub-keys, which are concatenated and privacy amplified so
that exposing a single sub-key does not significantly compromise the
final secret key.  This defence appears equally effective against
abort attacks on device-independent quantum randomness expansion
protocols.  Since quantum randomness expansion generally involves only
a single party, these protocols are not vulnerable to the impostor
attacks described in the main text.  It thus appears that it may be
possible in principle to completely defend them against memory
attacks, albeit at some cost.

It is also worth noting that there are many scenarios in which one
only needs short-lived randomness, for example, in many gambling
applications, bets are often placed about random data that are later
made public.  In such scenarios, once such random data have been 
revealed, the devices could be reused without our attacks presenting any
problem.

\end{document}